\documentclass[12pt]{article}
\usepackage{amsmath,amsthm}

\usepackage{array,multirow,graphicx}
\usepackage{float}
\usepackage{color}
\usepackage{graphicx,psfrag,epsf}
\usepackage{enumerate}
\usepackage{hyphenat}
\usepackage{natbib}
\usepackage{caption}
\usepackage{subcaption}
\usepackage{accents}

\usepackage{hyperref}
\setcitestyle{authoryear,open={(},close={)}}
\usepackage{authblk} 
\usepackage{url} 

\usepackage[dvipsnames]{xcolor}
\usepackage{tcolorbox}

\usepackage{mathrsfs}
\usepackage{enumitem}

\newcommand{\cadlag}{c\`{a}dl\`{a}g }
\usepackage[small]{titlesec}
\titleformat{\subsubsection}[runin]
{\normalfont\bfseries}{\thesubsubsection}{1em}{}
\usepackage{adjustbox}
\usepackage{prodint}
\usepackage{booktabs}
\usepackage{bm}
\usepackage{bbold}
\usepackage{wasysym}
\usepackage[margin=1.3in]{geometry}
\newcommand{\EE}{\mathbb{E}}

\newcommand{\eps}{\varepsilon}

\newcommand{\F}{\mathcal{F}}

\newcommand{\R}{\mathbb{R}}
\newcommand{\N}{\mathbb{N}}
\newcommand{\1}{\mathbb{1}}

\newcommand{\tmax}{\tau}

\newcommand{\argmin}{\text{argmin}}
\newcommand{\cupdot}{\mathbin{\mathaccent\cdot\cup}}

\newtheorem{thm}{Theorem}

\newtheorem{example}[thm]{Example}
\AtEndEnvironment{example}{\tiny$\!\!\bullet$\endexample}
\AtEndEnvironment{remark}{\tiny$\!\!\bullet$\endremark}

\newtheorem{remark}{Remark}
\newtheorem{assumption}{Assumption}

\usepackage{tikz}

\newcommand{\blind}{0} 

\date{\vspace{-1cm}}

\begin{document}

\def\spacingset#1{\renewcommand{\baselinestretch}%
{#1}\small\normalsize} \spacingset{1}


\if0\blind { \title{\bf Nonparametric efficient causal estimation of
    the intervention-specific expected number of recurrent events with
    continuous-time targeted maximum likelihood and highly adaptive
    lasso estimation } \author{
    Helene C. W. Rytgaard\\
    Section of Biostatistics, University of Copenhagen\\
    and \\
    Mark J. van der Laan \\
    Division of Biostatistics \\ and \\ Center for Targeted Machine
    Learning, University of California, Berkeley}
  \maketitle
} \fi

\if1\blind
{
    {\title{\bf Nonparametric efficient causal estimation of
    the intervention-specific expected number of recurrent events with
    continuous-time targeted maximum likelihood and highly adaptive
    lasso estimation
  }\author{}  \maketitle}
  \medskip
} \fi

\bigskip

\begin{abstract}

  Longitudinal settings involving outcome, competing risks and
  censoring events occurring and recurring in continuous time are
  common in medical research, but are often analyzed with methods that
  do not allow for taking post-baseline information into account. In
  this work, we define statistical and causal target parameters via
  the g-computation formula by carrying out interventions directly on
  the product integral representing the observed data distribution in
  a continuous-time counting process model framework. In recurrent
  events settings our target parameter identifies the expected number
  of recurrent events also in situations where the censoring mechanism
  depends on past information of post-baseline covariates such as the
  recurrent event process.  We propose a flexible estimation procedure
  based on targeted maximum likelihood estimation coupled with highly
  adaptive lasso estimation for double robust and nonparametric
  inference for the considered target parameter. We illustrate our
  methods in a simulation study, and an application to a real data
  example.

\end{abstract}

\noindent%

\vspace{0.1cm} \newpage

\spacingset{1.9} 

\section{Introduction}

Many problems of event history analysis and medical research involve
outcome and censoring events occurring and recurring in continuous
time. Recurrent events, specifically, refer to events that can occur
multiple times for an individual over the duration of a study. Some
examples include gastrointestinal infection in infants
\citep{schnitzer2014effect}, repeated hospital admissions
\citep{kessing2004predictive}, or recurring lung exacerbations in
cystic fibrosis patients \citep{miloslavsky2004recurrent}.  In this
work, we consider nonparametric efficient and double robust estimation
of a real-valued causal parameter in recurrent events settings
characterized by baseline confounding, right-censoring and competing
risks, particularly allowing for censoring mechanisms depending on the
past of the recurrent event process.

Historically, two primary approaches have characterized the analysis
of recurrent events \citep{cook2007statistical}, one based on
intensity models and another based on so-called marginal models.  In
intensity-based approaches
\citep{cook2002analysis,cook2007statistical}, models include different
variations over summaries of information on previous events via
time-dependent covariates
\citep{prentice1981regression,andersen1982cox,aalen2004dynamic,cook2007statistical},
and the effect measures targeted are defined as hazard ratios obtained
as regression coefficients. However, such hazard ratios do not attain
a meaningful interpretation, both because hazard ratios generally
suffer from lack of causal interpretability
\citep{hernan2010hazards,martinussen2020subtleties} and because
conditioning on post-treatment confounders blocks previous treatment
effects \citep{robins1986new}. Although different approaches have been
suggested to model the dependence among repeated events by modeling
intensities conditional on previous events via time-dependent
covariates \citep[see, e.g.,][]{andersen1982cox,aalen2004dynamic}, it
has generally been warned against conditioning on the number of past
events \citep{cook2007statistical} as these are critical cases of
post-randomization covariates.
While frailty models \citep{vaupel1979impact} provide methods to
incorporate heterogeneity without conditioning on the number of
recurrent events, the hazard ratios derived from these models maintain
only an individual-level interpretation and continue to lack causal
interpretability \citep[with certain model-specific exceptions for
which there is in fact agreement between the two,
see][]{hougaard2022choice}.  The other primary approach for analysis
of recurrent events instead directly target the expected number (the
marginal mean) of past recurrent events
\citep{lawless1995some,cook1997marginal,lin2000semiparametric},
estimation of which has for example been proposed carried out
nonparametrically \citep{cook1997marginal}, via regression modeling
\citep{ghosh2002marginal} or via pseudo-observations
\citep{andersen2019modeling,furberg2021bivariate}. In these proposals,
however, post-baseline information is generally not included, and the
often targeted nonparametric identification formula proposed by
\cite{cook1997marginal} will not coincide with the
(intervention-specific) expected number of recurrent events, involving
marginalization over the entire observed process, unless censoring is
independent. Particularly, such approaches do not allow
right-censoring depending on the past of the recurrent event process,
nor do they generalize to settings with time-dependent treatment
interventions.

Developments in the field of causal inference, on the other hand,
allow for correct adjustment for time-dependent confounders without
blocking the effect of previous treatments mediated through
time-dependent covariates \citep{robins1986new,robins1987addendum,
  robins1989analysis,robins1989control,
  robins1992estimation,robins1998marginal,robins2000marginal,robins2000robust,bang2005doubly,
  robins2008estimation,van2010targeted,van2010targetedII,petersen2014targeted}. Identification
of causal parameters can be done via inverse probability weighting,
such as done for recurrent events settings for example by
\cite{miloslavsky2004recurrent,janvin2024causal}, or, which is the
route we pursue in this work, via the g-computation formula
\citep{robins1986new,robins1987addendum}. The g-computation formula is
defined as some characteristic of the post-interventional distribution
obtained by carrying out hypothetical interventions directly on the
distribution of the data factorized according to the time-ordering. We
define our target parameter from the g-computation formula
corresponding to the intervention-specific expected number of
recurrent events, considering a counting process framework for the
observed data, where
the data-generating distribution is represented as a product of
factors involving product integrals
\citep{gill1990survey,andersen2012statistical}. For estimation, we
apply targeted maximum likelihood estimation
\citep{van2006targeted,van2011targeted} incorporating the efficient
influence curve \citep{bickel1993efficient} for the nonparametric
model, combined with highly adaptive lasso estimation to flexibly
account for dependence on, e.g., the follow-up information of the
recurrent event process. To the best of our knowledge, our methods are
the first to consider efficient nonparametric substitution estimation
for causal parameters in recurrent event settings; the closest related
work is \cite{cortese2022efficient} considering augmented inverse
probability weighted estimation of the marginal mean. In contrast to
their work, we target causal parameters under hypothetical treatment
interventions and we do not make the standard assumption of
independent censoring.  Additionally, we verify double robustness
properties, and present a different approach for nonparametric
estimation under weak conditions on the statistical model.  The recent
and still unpublished work by \cite{baer2023causal} considers
efficiency theory for a recurrent events setting as well. However,
their approach omits consideration of the historical information on
the recurrent event process, thereby limiting its ability to in fact
address efficiency and constraining their methodology's applicability
to real-world scenarios.

Our work extends and specializes the setting and methods considered by
\cite{rytgaard2021continuous} by allowing for recurrent events
outcomes; while the targeting procedures are closely related,
\cite{rytgaard2021continuous} proposes a targeting procedure based on
a combination of targeted maximum likelihood and sequential regression
update steps, whereas we here work directly with the g-computation
formula with direct plug-in estimation of clever covariates and the
target parameter, and our targeting procedure involves only estimation
and targeting of intensities. How to perform direct plug-in
estimation, which involves estimation of complex conditional
expectations across time, is an important topic of the present paper,
that will not only benefit the analysis of recurrent events data but
could potentially also be applied to enhance the estimation procedure
proposed by \cite{rytgaard2021continuous}.

There are two important components in the estimating procedure that we
propose, including 1) the estimation of conditional intensities and 2)
the algorithm to construct estimators for the conditional expectations
(clever covariates) entering the efficient influence curve, as well as
the target parameter, based on the estimators for the conditional
intensities. Regarding 1), which is widely recognized in the recurrent
events literature, it is a highly ambitious task to model the
distribution of the entire recurrent events process, with high risk of
model misspecification. In our approach we propose the use of highly
adaptive lasso estimation \citep{benkeser2016highly,van2017generally},
a nonparametric procedure which data-adaptively can learn about the
dependence on past information in the data without specification of
interaction or functional forms. While the highly adaptive lasso
estimator has been implemented for hazards of time-to-first-event
distributions \citep{rytgaard2021estimation}, with only baseline
information beyond survival status, we here consider specifically the
use of highly adaptive lasso estimator for intensities to learn about
the dependence on the past of the recurrent event process. Concerning
2), we describe a matrix-based recursive scheme for direct plug-in
evaluation, based on the interpretation of increments of
continuous-time counting processes as binary variables with
conditional probability distribution characterized by the
corresponding conditional intensity increment \citep[][Chapter
IV.1.5]{andersen2012statistical}. Essentially this involves replacing
all differentials by differences or jumps, and the resulting
estimation procedure is similar to how the classical Kaplan-Meier
estimator \citep{kaplan1958nonparametric} is formed as a finite
product over discrete increments of the Nelson-Aalen estimator, or the
Aalen-Johansen estimator \citep{aalen1978empirical} is formed as
finite integral over cause-specific hazard increments multiplied by
the Kaplan-Meier estimator.


The article is organized as follows.  Section
\ref{sec:statistical:estimation:setting} presents the statistical
estimation setting, including the observed data and the observed data
distribution. Section \ref{sec:target:parameter} defines the target
parameter.  The efficient influence curve is presented in Section
\ref{sec:nonparametric:eff:estimation}, as well as the second-order
remainder and its double robustness properties. {Section
  \ref{sec:collecting:inference} summarizes and presents results on
  nonparametric efficient inference for the presented targeted
  estimation procedure.}  Section \ref{sec:tmle} describes the
proposed targeted maximum likelihood estimation procedure updating
intensities for the recurrent event process and the death process.
Section \ref{sec:g:comp:::} describes estimation of conditional
expectations needed to evaluate the efficient influence curve and
estimate the target parameter, and flexible estimation via the highly
adaptive lasso of intensities depending on the history of the
recurrent event process.  Section \ref{sec:simulation:study} presents
a simulation study for proof of concept and as a demonstration of the
proposed methodology. Section \ref{sec:data:analysis} illustrates the
methods with a literature dataset concerning readmissions for
colorectal cancer patients.  Section \ref{sec:discussion} closes with
a discussion.

\section{Setting and notation}
\label{sec:statistical:estimation:setting}

We consider a recurrent events data setting
\citep{cook2007statistical,andersen2019modeling} as follows. Suppose
\(n\in\N\) subjects of population are followed in an interval of time
\([0,\tau]\), each with observed data characterized by a multivariate
counting process \citep{andersen2012statistical} \((N^y, N^d, N^c)\)
generating random times \({T}_1 < \cdots < {T}_{R(\tau)}\) at which a
recurrent outcome, the survival status and the censoring status may
change. For example, \(N^y\) could measure hospitalizations or number
of strokes across time. We assume that, at baseline, covariates
\(L\in \R^d\) are measured, and then a treatment decision
\(A\in\mathcal{A}\), taking value in a finite set \(\mathcal{A}\), is
made. We collect the observed data for any individual \(i\) in a
bounded interval \([0,t] \subseteq [0,\tau]\) as follows
\begin{align*}
  \bar{O}_i(t)  =\big( L_{i}, A_{i}, s,
  N_i^y(s), N_i^d(s) , N^c_i(s) \, :\, s \in\lbrace \tilde{T}_{i,r}\rbrace_{r=1}^{R_i(t)}\big), 
\end{align*}
where \(R_i(t)\) denotes the number of events experienced at time
\(t\). We let \(\F_{t} = \sigma( \bar{O}(t))\) denote the
\(\sigma\)-algebra generated by the observed data up until time \(t\),
and use \(\Lambda^{\cdot}\) to denote the cumulative intensity
corresponding to the counting process \(N^{\cdot}\), so that,
informally,
\( \EE_{P_0} [N^{\cdot}(dt)\mid \F_{t-}] = \Lambda_0^{\cdot}(dt \mid
\F_{t-})\). Thus, \(\Lambda_0^y\) denotes the non-terminal intensity
of the counting process tracking recurrent events, \(\Lambda_0^d\)
denotes the intensity of the terminal death process, and
\(\Lambda_0^c\) denotes the intensity of the terminal censoring
process.  Let \(T^d\) denote the survival time and \(T^c\) the
censoring time, so that we observe
\( T^{\mathrm{end}} = \min (T^d, T^c)\) and the indicator
\(\Delta = \1\lbrace T^d \le T^c\rbrace\). We further let
\(T^y_1 < T^y_2 < \cdots\) denote the times of occurrences for the
recurrent event. We thus have the observed times
\((T_1, \ldots, T_{R(\tau)}) = (T^y_1, \ldots, T^y_{R(\tau)-1},
T^{\mathrm{end}})\), where \(R(\tau)=1\) is allowed in which case
there is only one observation at \( T^{\mathrm{end}}\), and
\(T^y_1, \ldots, T^y_{R(\tau)-1} = \emptyset\). The observed counting
processes can also be represented as
\( N^d(t) = \1\lbrace T^{\mathrm{end}}\le t, \Delta=1\rbrace
\in\lbrace 0,1 \rbrace\),
\( N^c(t) = \1\lbrace T^{\mathrm{end}}\le t, \Delta=0\rbrace
\in\lbrace 0,1 \rbrace\) and
\( N^y(t) = \sum_{k} \1\lbrace T_k^y \le t, t \le T^{\mathrm{end}}
\rbrace \ge 0\), with \( t\in [0,\tau]\), \(\tau >0\).
We further denote by \(\mu_0\) the density of the distribution of
baseline covariates \(L\in \R^d\) with respect to a dominating measure
\(\nu\), and by \(\pi_0\) the distribution of the baseline treatment
decision \(A \in \mathcal{A}\), e.g.,
\(\mathcal{A} = \lbrace 0,1\rbrace\). The distribution \(P_0\) of the
observed data \(O= \bar{O}(\tau)\) across the entire interval
\([0,\tau]\) can then be represented generally as:
\begin{align}
  \begin{split}
  dP_0 (O) = \mu_{0}(L) d\nu(L) \pi_{0}(A \mid L)
  \qquad\qquad\qquad\qquad\qquad\qquad\qquad\qquad\qquad \\[0.2em]
  \Prodi_{s \le \tau} \big( \Lambda_0^y (ds \mid \F_{s-})\big)^{N^y(ds)}
  \big(1- \Lambda_0^y (ds \mid \F_{s-})\big)^{1-N^y(ds)} \\
  \Prodi_{s \le \tau} \big( \Lambda_0^d (ds \mid \F_{s-})\big)^{N^d(ds)}
  \big(1- \Lambda_0^d (ds \mid \F_{s-})\big)^{1-N^d(ds)} \\
  \Prodi_{s \le \tau} \big( \Lambda_0^c (ds \mid \F_{s-})\big)^{N^c(ds)}
  \big(1- \Lambda_0^c (ds \mid \F_{s-})\big)^{1-N^c(ds)} ,
  \end{split}\label{eq:factorize:P0}
\end{align}
with \(\prodi\) denoting the product integral \citep{gill1990survey}.
Note that the second factor
\( \prodi_{s \le \tau} ( 1- \Lambda_0^\cdot (ds \mid
\F_{s-}))^{1-N^\cdot(ds)}\) of each product integral evaluates to the
exponential form
\( \exp (-\int_0^\tau \Lambda_0^\cdot (ds \mid \F_{s-}))\) when
\( \Lambda_0^\cdot\) is continuous.

We further introduce notation for what we refer to as the
\textit{interventional}
\begin{align}
  dG_0 (O) &=  \pi_{0}(A \mid L)  \Prodi_{s \le \tau} \big( \Lambda_0^c (ds \mid \F_{s-})\big)^{N^c(ds)}
             \big(1- \Lambda_0^c (ds \mid \F_{s-})\big)^{1-N^c(ds)},\label{eq:G:part}
             \intertext{and \textit{non-interventional}}
             \begin{split} 
             dQ_0 (O) &=  \mu_{0}(L) d\nu(L)   \Prodi_{s \le \tau} \big( \Lambda_0^y (ds \mid \F_{s-})\big)^{N^y(ds)}
                        \big(1- \Lambda_0^y (ds \mid \F_{s-})\big)^{1-N^y(ds)}  \\
           & \qquad\qquad\qquad\qquad\qquad  \Prodi_{s \le \tau} \big( \Lambda_0^d (ds \mid \F_{s-})\big)^{N^d(ds)}
             \big(1- \Lambda_0^d (ds \mid \F_{s-})\big)^{1-N^d(ds)}, \end{split}\label{eq:Q:part}
\end{align}
parts of \(P_0\), i.e., \(dP_0=dQ_0 dG_0\). We will also use
\(dG_{0,s}, dQ_{0,s}\) to denote the \(s\)-specific factors. We assume
that \(P_0\) belongs to a statistical model \(\mathcal{M}\), where any
\(P\in\mathcal{M}\) can be parameterized as in \eqref{eq:factorize:P0}
and factorized as in \eqref{eq:G:part}--\eqref{eq:Q:part} into a
non-interventional part \(Q\) and an interventional part \(G\), making
only nonparametric assumptions for \(Q\) but allowing semiparametric
or parametric models for \(G\). We will also use the notation
\(P=P_{Q,G}\), and \(P_0=P_{Q_0,G_0}\) equivalently, to make the
parametrization into the non-interventional and interventional parts
explicit.

\section{Target parameter}
\label{sec:target:parameter}

We are generally interested in effects defined as parameters of the
g-computation formula \citep{robins1986new}, corresponding to
substituting a specific factor \(G\) of \(P\in\mathcal{M}\) by a
user-supplied (or estimated) choice \(G^*\). As specified in
\eqref{eq:G:part} above, \(G\) involves the treatment distribution
\(\pi\) as well as the censoring mechanism \(\Lambda^c\), and we shall
be interested in substituting \(G\) by \(G^*\) consisting of the
interventional \(\Lambda^{c,*} (dt \mid \F_{t-}) = 0\), putting all
mass in no right-censoring, and an interventional distribution
\(\pi^*(a \mid L) \) for the treatment decision \(A\) given covariates
\(L\). The latter could be, e.g., the degenerate distribution
\(\pi^*(a \mid L) = \1\lbrace a = a'\rbrace\) putting all mass in the
fixed value \(a'\in\mathcal{A}\).  We remark that any choice \(G^*\)
needs to satisfy the positivity assumption of absolute continuity,
i.e., that \(P^{G^*} \ll P\) for all \(P\in\mathcal{M}\), which means
for our purposes that
\begin{align}
  \frac{1-N^c(\tau-)}{\prodi_{t < \tau} ( 1- \Lambda^c(dt \mid \F_{t-}))}
  \frac{\pi^*(a \mid L)}{\pi(a \mid L)} < \infty,  \quad P-\text{a.s.} \text{ for } a=0,1.
  \label{eq:ass:positivity}
\end{align}
Note that this may also confine the choice of time horizon at which
the target parameter is evaluated{, as \eqref{eq:ass:positivity}
  implies that some individuals must remain uncensored by time
  \(\tau\) for any possible observed level of history}.

For a given choice \(G^*\), we then define our target parameter
\(\Psi \, :\, \mathcal{M} \rightarrow \R_+\) as the
intervention-specific mean outcome of the number of recurrent events
by time \(\tau >0\), i.e.,
\begin{align}
  \Psi(P) = \EE_{P^{G^*}} [N^y(\tau)],
  \label{eq:def:target:parameter}
\end{align}
where \(\EE_{P^{G^*}}\) denotes the expectation with respect to the
g-computation formula \(P^{G^*} = P_{Q,G^*}\).  The targeted causal
interpretation of \eqref{eq:def:target:parameter} is that it
represents the expected number of recurrent events if everyone had
been treated according to the distribution \(\pi^*\) and had there
been no censoring.  This holds under coarsening at random (CAR) on
censoring with respect to the full data structure without
right-censoring, and under conditional exchangeability with respect to
treatment in a world without censoring.  We address this more
thoroughly in the Supplementary Material.  We stress that the
censoring mechanism is allowed to depend on the observed parts of the
history, which contrasts with much of the recurrent events literature,
where a stricter independent censoring assumption is commonly used
\citep{cook1997marginal,cortese2022efficient,baer2023causal},
requiring that censoring is independent of the death and the recurrent
event processes, either marginally or only dependent on baseline
information alone.

\section{Nonparametric efficient estimation}
\label{sec:nonparametric:eff:estimation}

The efficient influence curve
\citep{bickel1993efficient,van2000asymptotic} for
\(\Psi \, : \, \mathcal{M}\rightarrow \R_+\) under the nonparametric
model \(\mathcal{M}\) tells us about conditions under which we may
construct \(\sqrt{n}\)-consistent and asymptotically linear
estimators, also when data-adaptive estimators are employed for
modeling the data-generating distribution through nuisance
parameters. This relies on nuisance parameter estimators constructed
such as to solve the efficient influence curve equation and to achieve
asymptotic negligibility of an empirical process term and of a
remainder term \citep[][Theorem
A.5]{van2006targeted,van2011targeted}. The basis is the so-called von
Mises expansion of the functional
\(\Psi \, : \, \mathcal{M}\rightarrow \R_+\),
\(\Psi (P) - \Psi(P') = \int_{\mathcal{O}} \phi^* (P) (o) d(P-P')(o) +
R(P,P')\), where \(\phi^*(P)\) denotes the efficient influence curve,
\(R(P,P') = \Psi(P) - \Psi(P') + P' \phi^*(P)\) denotes the mentioned
remainder term, and \(P,P'\in\mathcal{M}\) are distributions in the
statistical model \(\mathcal{M}\). Note that \(dP\) denotes
integration with respect to the probability measure \(P\) for the
random variable \(O\); we will also use the common operator notation
\(P h = \int h(o) dP(o)\) for functions
\(h \, : \, \mathcal{O} \rightarrow \R\), and equivalently write
\( P h = \EE_{P} [h(O)]\).
In Section \ref{sec:efficient:influence} we first present the
efficient influence curve for our target parameter, in Section
\ref{sec:second:order:remainder} we next present double robust
properties, and in Section \ref{sec:collecting:inference} we collect
results and conditions for obtaining inference based on the efficient
influence curve for our proposed estimator.

\subsection{Efficient influence curve}
\label{sec:efficient:influence}

We here present the efficient influence curve for estimating
\(\Psi \, :\, \mathcal{M} \rightarrow \R_+\) in the nonparametric
model \(\mathcal{M}\).
With the notation
\(\Delta N^{\cdot}(t) = N^{\cdot}(t) - N^{\cdot}(t-)\), we define
so-called clever covariates as follows
  \begin{align}
    h^d_t (\Lambda^d,\Lambda^y) (O) &=  N^y(t-) - \EE_{P^{G^*}} \big[ N^y (\tau) \, \big\vert\,
                                      \Delta N^d(t)=0, \F_{t-}\big],
                          \label{eq:clever:d}\\
    \begin{split}
      h^y_t  (\Lambda^d,\Lambda^y) (O)  &  =  \EE_{P^{G^*}} \big[ N^y (\tau) \, \big\vert\, \Delta N^y(t) = 1, \F_{t-}\big] 
                                        - \EE_{P^{G^*}} \big[ N^y (\tau) \, \big\vert\,\Delta N^y(t) = 0,  \F_{t-}\big]  ,
    \end{split}    \label{eq:clever:y}
    \intertext{and further  clever weights as}
    w_t (\pi, \Lambda^c) (O) &= \frac{\pi^* (A\mid L) }{\pi (A\mid L) \prodi_{s < t} (1- \Lambda^c(ds \mid  \F_{s-}))}. 
                               \label{eq:clever:weight}
  \end{align}
  Then the efficient influence curve is given by
  \begin{align}
  &\phi^* (P) (O)
  =
    \int_{t\le\tau} \1\lbrace T^{\mathrm{end}} \ge t\rbrace
    w_t (\pi, \Lambda^c) (O) h^d_t(\Lambda^d,\Lambda^y)(O)
    \big( N^d(dt) - \Lambda^d(dt \mid \F_{t-})\big)         \label{eq:effi:if:d}
  \\
  & \qquad + \,\int_{t\le\tau} \1\lbrace T^{\mathrm{end}} \ge t\rbrace
    w_t (\pi, \Lambda^c) (O)    h^y_t  (\Lambda^d,\Lambda^y) (O)  \big( N^y(dt) - \Lambda^y(dt\mid  \F_{t-})\big)
            \label{eq:effi:if:y} \\
  & \qquad + \,   \EE_{P^{G^*}} [ N^y(\tau) \mid L] - \Psi(P).         \label{eq:effi:if:mu}
\end{align}
The derivation can be found in the Supplementary Material.

\subsubsection{Remark on the stricter independent censoring assumption.}
  \label{rem:recurrent:events:no:dependence}

  Consider the case that censoring is independent of the death and the
  recurrent event processes, either marginally or only dependent on
  baseline information.  As remarked upon in Section
  \ref{sec:target:parameter}, this is a common assumption in the
  analysis of recurrent events.  Denote the corresponding statistical
  model by \(\tilde{\mathcal{M}} \subset\mathcal{M}\). On the smaller
  model \(\tilde{\mathcal{M}}\), a valid representation for the
    target parameter is
 \begin{align}
   \tilde{\Psi}(P) =  \int_{\R^d} \sum_{a=0,1} \bigg(\int_0^{\tau} \tilde{S}^d(t- \mid  a,\ell)
   \tilde{\Lambda}^y(dt \mid  T^{\mathrm{end}} \ge t, a, \ell)\bigg)
    \pi^*(a \mid \ell) d\mu (\ell),
    \label{eq:simple:nonparametric}
  \end{align}
  with
  \(\tilde{\Lambda}^y(dt \mid T^{\mathrm{end}} \ge t, A, L ) = \EE_P [
  N^y (dt) \mid T^{\mathrm{end}} \ge t, A, L ]\) denoting the
  recurrent event rate (not intensity), and
  \(\tilde{S}^d(t \mid a,\ell) =\prodi_{s \in (0, t]} (1 -
  \Lambda^d(ds \mid T^{\mathrm{end}} \ge s, a,\ell)) \) denoting the
  survival function based on the death rate (not intensity)
  \(\tilde{\Lambda}^d(dt \mid T^{\mathrm{end}} \ge t, A, L ) = \EE_P [
  N^d (dt) \mid T^{\mathrm{end}} \ge t, A, L ]\).
  Note that plugging the Kaplan-Meier estimator for \(\tilde{S}^d\)
  and the Nelson-Aalen estimator for \(\tilde{\Lambda}^y\) (both
  stratified on \(A\)) into \eqref{eq:simple:nonparametric}
  corresponds to a stratified version of the nonparametric estimator
  initially proposed by \cite{cook1997marginal}.
  
  However, an estimator based directly on
  \eqref{eq:simple:nonparametric}, not taking into account follow-up
  information on the recurrent event process, can only be locally
  efficient.  This could for example be a TMLE based on working models
  \(\tilde{\Lambda}^d,\tilde{\Lambda}^y\) for \(\Lambda^d,\Lambda^y\)
  leaving out follow-up information on the recurrent process. With
  such working models, the clever covariates from Equation
  \eqref{eq:clever:d} and \eqref{eq:clever:y} reduce to
  \( \tilde{h}^y_t (\Lambda^d,\Lambda^y) (O) = 1 \) and
  \( \tilde{h}^d_t (\Lambda^d,\Lambda^y) (O) = -\frac{\int_t^{\tau}
    \tilde{S}^d(s- \mid A,L) \tilde{\Lambda}^y(ds \mid
    T^{\mathrm{end}} \ge s, A,L) }{ \tilde{S}^d (t \mid A,L)}\), and
  the working model based TMLE will agree with estimators starting
  with \eqref{eq:simple:nonparametric} rather than the full
  g-computation formula representation
  \eqref{eq:def:target:parameter}.

\subsection{Double robustness properties}
\label{sec:second:order:remainder}

The remainder from the von Mises expansion, defined as
\(R(P, P_0) = \Psi(P) - \Psi(P_0) + P_0 \phi^*(P)\), plays a crucial
role in understanding the double robustness properties of the
estimation problem. In the Supplementary Material we show that
\begin{align*}
   R(P,P_0)   =
  \sum_{x=y,d}  \EE_{P^{G^*}_0}\bigg[ \int_0^{\tmax} 
    \bigg(  \frac{\pi_0 (A\mid L)   \prodi_{s < t} (1-\Lambda_0^c(ds \mid \F_{s-})) }{\pi (A\mid L) \prodi_{s < t} (1- \Lambda^c(ds \mid  \F_{s-}))} - 1\bigg)
  (1-N^d(t-))
  \\[-0.2cm]
    \times \,
    h_t^x(\Lambda^d, \Lambda^y) (O) \big( \Lambda^x_0( dt \mid \F_{t-}) - \Lambda^x( dt \mid \F_{t-}) \big) \bigg].
\end{align*}
This displays an important double robust structure, with direct
implications for:

\textit{1. Double robustness in consistency}. We have that
\(R(P,P_0) = 0\) if either \(\pi_0(a\mid \ell) = \pi(a \mid \ell)\)
and
\(\prodi_{s<t} \Lambda^c_0(ds \mid \F_{s-}) = \prodi_{s<t}
\Lambda^c(ds \mid \F_{s-})\), or if
\(\Lambda^d_0 (dt \mid \F_{t-} ) = \Lambda^d (dt \mid \F_{t-} )\) and
\(\Lambda^y_0 (dt \mid \F_{t-} ) = \Lambda^y (dt \mid \F_{t-} )\) for
all \(t \le \tau\). Thus, consistent estimation can be achieved if
consistent estimators are provided for either the censoring
distribution and the treatment distribution, or for the intensities of
the recurrent event and the death processes.
  
\textit{2. Rate double robustness}. The double robust structure
  implies that the convergence rate of the second-order remainder can
  be bounded by a product of rates for the non-interventional and for
  the interventional parts, respectively. This implies, for example,
  that we have that \(R (\hat{P }^*_n , P _0 ) = o_P(n^{-1/2})\) for
  an estimator \(\hat{P }^*_n\) when \(\hat{P }^*_n\) consists of
  estimators for the nuisance parameters
  \((\Lambda^y, \Lambda^d, \Lambda^c, \pi)\) each estimating their
  true counterpart at a rate faster than
  \(o_P(n^{-1/4})\).

\subsection{Inference for the targeted estimator } 
\label{sec:collecting:inference}

Section \ref{sec:efficient:influence} provides an expression for the
efficient influence curve depending on the data-generating
distribution through the nuisance parameters
\((\Lambda^y, \Lambda^d, \Lambda^c, \pi)\). In Section \ref{sec:tmle}
we use this directly to construct a targeted fluctuation procedure to
update a set of initially given estimators for the nuisance parameters
\(\hat{P}_n = (\hat{\Lambda}_n^y, \hat{\Lambda}_n^d,
\hat{\Lambda}_n^c, \hat{\pi}_n) \mapsto \hat{P}^{\mathrm{tmle}}_n \)
such as to solve the efficient influence curve equation,
\begin{align}
  \mathbb{P}_n \phi^* ( \hat{P}^{\mathrm{tmle}}_n ) = o_P(n^{-1/2}).
  \label{eq:eic:equation}
\end{align}
Note that \(\mathbb{P}_n\) denotes the empirical distribution, such
that
\(\mathbb{P}_n \phi^* ( \hat{P}^{\mathrm{tmle}}_n ) = \tfrac{1}{n}
\sum_{i=1}^n \phi^* ( \hat{P}^{\mathrm{tmle}}_n )(O_i)\). The targeted
fluctuation procedure essentially amounts to a bias correction step,
and is an important part of establishing asymptotic linearity. The
proof \citep[see, e.g.,][Theorem
A.5]{van2006targeted,van2017generally,rytgaard2021continuous,van2011targeted}
consists in evaluating the von Mises expansion in the estimator
\( \hat{P}^{\mathrm{tmle}}_n\) and the true data-generating \(P_0\),
and expanding this into a sum including multiple terms that need to be
asymptotic negligible, one corresponding to the efficient influence
curve equation, one to an empirical process term, and one to the
second-order remainder. Combined with highly adaptive lasso
estimation, we need the following conditions to hold. We emphasize, as
we will return to later, that these conditions impose much weaker
assumptions than what is needed to rely on (semi)parametric modeling.

\begin{assumption}
  For any \(P\in\mathcal{M}\) the nuisance parameters
  \((\Lambda^y, \Lambda^d, \Lambda^c, \pi)\) can be parametrized by
  \cadlag (i.e., right-continuous with left limits) functions with
  finite sectional variation norm, and \( (\Lambda^c, \pi)\) together
  with the choice \(\tau >0\) further fulfill the (strong) positivity
  assumption that
  \(\prodi_{t < \tau} ( 1- \Lambda^c(dt \mid \F_{t-})) \pi(a \mid L) >
  \eta > 0 \), a.e., for \(a=0,1\).
  \label{ass:conditions:asymptotic:linearity}
\end{assumption}

Under Assumption \ref{ass:conditions:asymptotic:linearity} it holds
for an estimator \( \hat{P}^{\mathrm{tmle}}_n\) fulfilling
\eqref{eq:eic:equation}, when the estimators for the nuisance
parameters are constructed such as to achieve at least \(n^{-1/4}\)
convergence (such as with highly adaptive lasso estimation), that the
TMLE estimator
\(\hat{\psi}^{\mathrm{tmle}}_n = \Psi (\hat{P}_n^{\mathrm{tmle}})\)
for the target parameter is asymptotically linear with influence
function equal to the efficient influence curve, 
\(\sqrt{n} ( \hat{\psi}^{\mathrm{tmle}}_n - \Psi ( P_0 ) ) = \sqrt{n}
\, \mathbb{P}_n \phi^*( P_0) + o_P(n^{-1/2})\). From this it directly
follows that
\( \sqrt{n} ( \Psi (\hat{P}_n^{\mathrm{tmle}}) - \Psi ( P_0 ) )
\overset{\mathcal{D}}{\rightarrow} N(0, P_0 \phi^*(P_0)^2)\), and
inference for \(\hat{\psi}^{\mathrm{tmle}}_n \) follows.

\section{Targeted maximum likelihood estimation}
\label{sec:tmle}

Targeted maximum likelihood proceeds by fluctuating initial estimators
by maximizing the likelihood in a targeted direction such as to solve
the efficient influence curve equation. In our case, the entire
data-generating distribution is parameterized by intensities, and we
can carry out targeting according to the log-likelihood loss function
and least favorable submodel presented by \citet[][Section
5.2]{rytgaard2021continuous}, which was similarly applied for
competing risks settings in \cite{rytgaard2021estimation}. 

For the recurrent event specific and the death specific cumulative
intensities \(\Lambda^y, \Lambda^d\) with corresponding intensity
processes \(\lambda^y, \lambda^d\), we define the intercept-only
submodel (\(x=y,d\))
\begin{align}
  \Lambda_{\eps}^x (dt\mid \F_{t-}) =
  \Lambda^x (dt\mid \F_{t-}) \exp( \eps), \quad \eps\in\R,
  \label{eq:Lambda:submodel}
\end{align}
and further the log-likelihood loss function
\begin{align*}
  \mathscr{L}_x(\Lambda^x ) (O) = \int_0^{\tau} w_t (\pi, \Lambda^c) (O) (O) h^x_t (\Lambda^d,\Lambda^y)(O) \big( \log \lambda^x(t \mid \F_{t-}) N^x(dt)
  - \Lambda^x (dt \mid \F_{t-})\big). 
\end{align*}
It is straightforward to show that this pair has the desired property
that
\begin{align*}
  \frac{d}{d\eps}\bigg\vert_{\eps = 0}  \mathscr{L}_x(\Lambda_\eps^x ) (O) =
  \int_0^{\tau}  w_t (\pi, \Lambda^c) (O) (O) h^x_t (\Lambda^d,\Lambda^y)(O) \big( N^x(dt) -  \Lambda^x (dt \mid \F_{t-})\big). 
\end{align*}
We note that alternative versions are also possible, where the
\textit{clever covariate} and/or the \textit{clever weight}, are
included as a (clever) covariate in the fluctuation model in the
submodel, rather than used as (clever) weights in the loss function.

\subsection{Targeting procedure}
\label{sec:special:tmle}

Our targeting procedure consists of the following:

\textit{1. Initial estimation.}  Estimators \(\hat{\Lambda}_n^c\) and
\(\hat{\pi}_n\) for \(\Lambda^c\) and \(\pi\), based on which we get
an estimator for \(w_t(\pi,\Lambda^c)\) as:
  \begin{align*}
    w_t (\hat{\pi}_n,\hat{\Lambda}_n^c)(O) = 
\frac{\pi^* (A\mid L) }{\hat{\pi}_n (A\mid L) \prodi_{s < t} (1- \hat{\Lambda}_n^c(ds \mid  \F_{t-}))}    . 
  \end{align*}
Furthermore, initial estimators \(\hat{\Lambda}_{n}^d\) and
  \(\hat{\Lambda}_{n}^y\) for \(\Lambda^d\) and \(\Lambda^y\), based
  on which we also estimate \( h^d_t (\Lambda^d,\Lambda^y)\) and
  \( h^y_t (\Lambda^d,\Lambda^y) \); while the latter is
  straightforward if estimators \(\hat{\Lambda}_{n}^d\) and
  \(\hat{\Lambda}_{n}^y\) are based on working models leaving out
  follow-up information on the recurrent process (see Section
  \ref{rem:recurrent:events:no:dependence}), it is much more subtle
  for the general case. This is the topic of Section
  \ref{sec:g:comp:::}.

\textit{2. Targeting.} A targeting procedure to update \(\Lambda^d\) and
    \(\Lambda^y\). We propose to execute this in an iterative manner
    starting with \(\hat{\Lambda}_{n,0}^{d} := \hat{\Lambda}_{n}^d\)
    and \(\hat{\Lambda}_{n,0}^{y} := \hat{\Lambda}_{n}^y\), and the
    \(m\)th step proceeding as follows:
\begin{description}
\item[Update
  \(\hat{\Lambda}_{n,m}^{d} \mapsto \hat{\Lambda}_{n,m+1}^{d}\)] along
  the parametric submodel \eqref{eq:Lambda:submodel} to solve the
  relevant term \eqref{eq:effi:if:d} equal to zero for given
  \(w_t(\hat{\Lambda}_n^c, \hat{\pi}_n)\),
  \( h^d_t (\hat{\Lambda}_{n,m}^{d}, \hat{\Lambda}_{n,m}^{y})\) and
  \( h^y_t (\hat{\Lambda}_{n,m}^{d}, \hat{\Lambda}_{n,m}^{y}) \).
\item[Update
  \(\hat{\Lambda}_{n,m}^{y} \mapsto \hat{\Lambda}_{n,m+1}^{y}\)] along
  the parametric submodel \eqref{eq:Lambda:submodel} to solve the
  relevant term \eqref{eq:effi:if:y} equal to zero for given
  \(w_t(\hat{\Lambda}_n^c, \hat{\pi}_n)\),
  \( h^d_t (\hat{\Lambda}_{n,m}^{d}, \hat{\Lambda}_{n,m}^{y})\) and
  \( h^y_t (\hat{\Lambda}_{n,m}^{d}, \hat{\Lambda}_{n,m}^{y}) \).
\end{description}
This procedure is repeated until
\(\vert \, \mathbb{P}_n \phi(\hat{P}_n^{*}) \, \vert \le s_n\), where
\(\hat{P}_n^{*}= (\hat{\Lambda}^{y(m^*)}_{n},
\hat{\Lambda}^{d(m^*)}_{n},\hat{\Lambda}^c_n, \hat{\pi}_n)\) and
\(s_n = \sqrt{\mathbb{P}_n (\phi ( \hat{P}_n))^2} / (n^{-1/2} \log
n)\) with \(\mathbb{P}_n ( \phi ( \hat{P}_n))^2\) estimating the
variance of the efficient influence function based on the collection
of initial estimators for the nuisance parameters
\(\hat{P}_n= (\hat{\Lambda}^y_{n},
\hat{\Lambda}^d_{n},\hat{\Lambda}^c_n, \hat{\pi}_n)\).

\section{Clever covariate estimation}
\label{sec:g:comp:::}

The clever covariates from Equations \eqref{eq:clever:d} and
\eqref{eq:clever:y} presented in Section \ref{sec:efficient:influence}
require evaluation of the conditional expectations
\(\EE_{P^{G^*}} [ N^y (\tau) \, \vert\, \Delta N^d(t) = 0,
T^{\mathrm{end}} \ge t, \F_{t-}]\),
\(\EE_{P^{G^*}} [ N^y (\tau) \, \vert\, \Delta N^y(t) = 1,
T^{\mathrm{end}} \ge t, \F_{t-}]\) and
\(\EE_{P^{G^*}} [ N^y (\tau) \, \vert\, \Delta N^y(t) = 0,
T^{\mathrm{end}} \ge t, \F_{t-}]\) under the g-computation formula. In
Section \ref{sec:clever:covar:g:comp}, we describe a plug-in approach
based on a matrix-based recursion scheme for direct evaluation of the
g-computation formula; in Section \ref{sec:highly:adaptive:lasso} we
then describe the incorporation, and particular utility, of highly
adaptive lasso estimation for estimation of the conditional
intensities.

We shall overall move forward with Andersen-Gill type multiplicative
models \citep{andersen1982cox,andersen2012statistical} for \(x=y, d, c\) as follows:
\begin{align}
  \begin{split}
  \Lambda^x (dt \mid \F_{t-}) =
  \lambda^x (t \mid \F_{t-})dt & = 
                                 \1\lbrace  T^{\mathrm{end}} \ge t\rbrace
                                 \lambda^x_{\mathrm{bl}} (t) \exp ( f^x(t, \rho(\bar{N}^y(t-)), A,L)) dt ,
                             \end{split}      \label{eq:ag:multi:models}
\end{align}
where
\( \lambda^y_{\mathrm{bl}}, \lambda^d_{\mathrm{bl}},
\lambda^c_{\mathrm{bl}}\) are unspecified baseline hazards, and
\(f^y, f^d, f^c\) are functions of time \(t\), treatment \(A\) and
covariates \(L\) as well as the past of the recurrent event process
\(\bar{N}^y(t-) = (N^y(s) \, : \, s <t)\) via a summary function
\(\rho\).  We remark that the baseline intensities may also be
collapsed into \(f^y\), \(f^d\) and \(f^c\). The function \(\rho\)
expresses what kind of summaries of \(\bar{N}^y(t-)\) that
\(f^y,f^d, f^c\) may depend on, whereas \(f^y, f^d, f^c\) specify
functional forms and interactions with time, treatment or
covariates. We assume that \(\rho\) is fixed a priori, while \(f^y\),
\(f^d\) and \(f^c\) may be data-adaptively learned from the data,
e.g., by highly adaptive lasso estimation as we return to in Section
\ref{sec:highly:adaptive:lasso}.
To reflect the potential data-adaptivity, we will use the notation
\(f^{\cdot}_n\).  Note also that \(\rho\) need not be a fixed model
assumption, and that multiple HAL estimators with different choices of
\(\rho\) are possible.

\subsection{Analytic evaluation of the g-computation formula}
\label{sec:clever:covar:g:comp}

Throughout this section, we denote the ordered observed event times
across all subjects by
\begin{align*}
  0 = T_{(0)} < T_{(1)} <\cdots < T_{(K)} \le \tau, 
\end{align*}
and we use the shorthand notation
\(\bar{O}_{(k)} = \bar{O}({T_{(k)}})\),
\(\F_{k} = \sigma(\bar{O}_{(k)})\) and
\(\F_{k-} = \sigma(\bar{O} (T_{(k)}-))\), for \(k=1, \ldots,K\).
For the following, we consider discrete cumulative intensities
\(\Lambda^{\cdot}\) that have jumps where \(N^{\cdot}\) does. The
distribution of increments \( \Delta N^{\cdot} ( T_{(k)} )\) given the
past up till time \(T_{(k)}\) can be characterized by
\( \Delta \Lambda^{\cdot} ( T_{(k)} \mid \F_{k-}) \), and we shall
construct an estimation procedure for the clever covariates based on
sequentially integrating out the `binary variables' corresponding to
increments \( \Delta N^{\cdot} ( T_{(k)} )\). In the Supplementary
Material we exemplify how this process yields well-known formulas in
the point-treatment survival and competing risks settings, as well as
a recurrent event setting where intensities do not depend on the
history of the recurrent event process. 

We emphasize that the discreteness of the cumulative intensities
\(\Lambda^{\cdot}\) does not reflect an assumption of the parameter
space, but is used rather to reflect the discreteness of
estimators. For example, when \(\Lambda^{\cdot}\) is estimated by a
Nelson-Aalen estimator \citep{johansen1978product}, it will indeed be
a step function with increments at the observed jump times. Likewise
this is the case when estimation is based on a Cox model combined with
a Breslow estimator for the cumulative baseline hazard
\citep{breslow1974covariance}. When we proceed to highly adaptive
lasso estimation in Section \ref{sec:highly:adaptive:lasso}, with the
implementation based on a penalized Poisson regression approach, the
increments of cumulative intensities over intervals
\((T_{(k)}, T_{(k+1)}]\) are obtained by multiplying the intensity
rate at time \(T_{(k)}\) by the length of the interval
\((T_{(k)}, T_{(k+1)}]\).

\subsubsection{Piecewise constant dependence on the recurrent event
  process}
\label{sec:clever:notation:index}

The crucial part for developing the following estimators for the
clever covariates comes down to the number of (finitely many)
different possible paths that we need to integrate out over. This is
determined by \(f^{\cdot}_n\), and what we ultimately need is that
\( f^{\cdot}_n(t, \rho(\bar{N}^y(t-)), a,\ell)\) varies as a piecewise
constant function of its argument \(\bar{N}^y(t-)\). Then we can
partition the sample space of \(\bar{N}^y(t-)\) into cubes
\(\cupdot_{j\in\mathscr{J}} \mathcal{N}_j\) for a finite index set
\(\mathscr{J}\), such that
\begin{align}
  {f^{\cdot}_n}(t, \rho(\bar{N}_1^y(t-)), a,\ell)
  =  {f^{\cdot}_n}(t, \rho(\bar{N}_2^y(t-)), a,\ell), \quad \text{for all }
  \bar{N}_1^y(t-), \bar{N}_2^y(t-) \in \mathcal{N}_j,
  \label{eq:f:n:dependence:grid}
\end{align}
for fixed \(t, a, \ell, j\). Note that \({f^{\cdot}_n}\) is
allowed to vary more generally in its remaining arguments.

There are different ways to achieve this piecewise constant
dependence. It is for example ensured for any \({f_n^{\cdot}}\)
when \(\rho\) is simply fixed beforehand so that it maps
\(\bar{N}^y(t-)\) to simple categorical variables, such as, for
example, \(\rho (\bar{N}^y(t-) ) = \1 \lbrace N^y(t-) \ge 1\rbrace\)
(any events in the past), or \(\rho (\bar{N}^y(t-) ) = N^y(t-) \)
(number of events in the past). More generally it is ensured when
\({f^{\cdot}_n}\) is obtained with the highly adaptive lasso
estimator, which by construction depends on covariates through a
linear combination of zero-order spline basis functions.
Then, even if \(\rho (\bar{N}^y(t-) ) \) should depend more generally
on \( \bar{N}^y(t-) \), there will still be a partitioning
\(\cupdot_{j\in\mathscr{J}} \mathcal{N}_j\) of the sample space of
\(\bar{N}^y(t-)\) such that \eqref{eq:f:n:dependence:grid} holds.

\subsubsection{Evaluation of the g-computation formula and clever covariates} 
\label{sec:evaluating:g:computation}

We shall here use \(\Delta\Lambda_{k,j'}^{\cdot}(a,L)\) to denote the
time \(T_{(k)}\)-increment of the intensity \(\Lambda^{\cdot}\)
evaluated in the histories up till time \(T_{(k-1)}\) that are
consistent with index \(j\), i.e.,
\begin{align*}
  \Delta\Lambda^{\cdot}_{k,j} (a,L) = 
  \Delta \Lambda^{\cdot} ( T_{(k)} \mid T^{\mathrm{end}} \ge T_{(k)},  \bar{N}^y( T_{(k-1)})
{\,\in \mathcal{N}_j}, A=a, L). 
\end{align*}
%
%
Similarly, we define objects
\begin{align}
  Z_{k+1,j} (a , L)
  & = \EE_{P^{G^*}} [ N^y(\tau) - N^y (T_{(k+1)}) \mid  \bar{N}^y (T_{(k)})  {\,\in \mathcal{N}_{j}}  ,
    T^{\mathrm{end}} \ge T_{(k{+1})}
    , A=a, L],
    \label{eq:Z:k+1}
\end{align}
which are the conditional expectations under the g-computation formula
with the conditioned history prior to time \(T_{(k)}\) being
consistent with index \(j\). Note that we have
\(Z_{K-1,j} (a , L) = \Delta \Lambda^y_{K,j} (a, L)\), and that
\(Z_{0,j} (a , L) = \EE_{P^{G^*}} [ N^y(\tau) \mid A=a, L]\)
irrespective of \(j\).  We here focus the presentation on the case
that \(\rho(\bar{N}^y(t-)) = \min ( N^y(t-), J-1)\), for which
\(j\in \lbrace 1,2,\ldots, J\rbrace\) marks that \( N^y(t-)= j-1\).
The presentation for general \(\rho(\bar{N}^y(t-))\) can be found in
the Supplementary Material together with the proof of the following
procedure for evaluating the conditional expectations
\(\bm{Z}_k (a,L) = (Z_{k,j} (a,L)\, : \, j \in \mathscr{J})\in
\R^{J\times 1}\) from \(\bm{Z}_{k+1}(a,L)\).  Denote by
\(\bm{Z}_k (a,L) = (Z_{k,j} (a,L)\, : \, j \in \mathscr{J})\in
\R^{J\times 1}\), and define the augmented
\(\tilde{\bm{Z}}_k (a,L) =(\bm{Z}_k (a,L), 1)\in \R^{(J+1)\times
  1}\). For all \(a,L\), we have that:
\begin{align*}
  \tilde{\bm{Z}}_k (a,L) =  \tilde{\bm{B}}_k(a,L) \tilde{\bm{Z}}_{k+1}(a,L) , 
\end{align*}
for a matrix \(\tilde{\bm{B}}_k (a,L) \in \R^{(J+1) \times (J+1)}\)
with elements (suppressing the dependence on \(a,L\)), for
\(j=1,\ldots, J-1\),
\(\tilde{\bm{B}}_{k,j,j} = (1- \Delta \Lambda^d_{k,j} )( 1-
\Delta\Lambda^y_{k,j} ) \),
\(\tilde{\bm{B}}_{k,j,j+1} = \tilde{\bm{B}}_{k,j,J+1}= \tilde{\bm{B}}_{k,J,J+1} = (1- \Delta
\Lambda^d_{k,j} )\Delta\Lambda^y_{k,j}  \),  for
\(j=J\),
\(\tilde{\bm{B}}_{k,j,j} = (1- \Delta \Lambda^d_{k,j} ) \), and with
\(\tilde{\bm{B}}_{k,J+1,J+1} =1\), and \(\tilde{\bm{B}}_{k,j,l} =0\)
otherwise.  
We also have that:
\begin{align*}
  h^d_{T_{(k)}}  (\Lambda^d,\Lambda^y) (O)  = \tilde{\bm{b}}^d_k(a,L)  \tilde{\bm{Z}}_{k+1}(a,L),
\end{align*}
where \(\tilde{\bm{b}}^d_k(a,L) \in \R^{1\times (J+1)}\) has all
entries equal to zero except
\(\tilde{\bm{b}}^d_{k,J+1} =
-\Delta\Lambda^{y}_{k,j(\bar{O}_{(k-1)})}\), and, for
\(j(\bar{O}_{(k-1)}) \le J-1\),
\(\tilde{\bm{b}}^d_{k,j(\bar{O}_{(k-1)})+1} = -
\Delta\Lambda^{y}_{k,j(\bar{O}_{(k-1)})}\) and
\(\tilde{\bm{b}}^d_{k,j(\bar{O}_{(k-1)})} = -(1-
\Delta\Lambda^{y}_{k,j(\bar{O}_{(k-1)})})\), and, for
\(j(\bar{O}_{(k-1)})=J\), \(\tilde{\bm{b}}^d_{k,J} = -1\).
Here  \(j(\bar{O}_{(k-1)})\) is the index consistent the observed
history \(\bar{O}_{(k-1)}\) of \(O\).
Likewise we have:
\begin{align*}
  h^y_{T_{(k)}}  (\Lambda^d,\Lambda^y) (O)  =  
  \tilde{\bm{b}}^y_k(a,L)  \tilde{\bm{Z}}_{k+1}(a,L),
\end{align*}
where \(\tilde{\bm{b}}^y_k(a,L) \in \R^{1\times (J+1)}\) has all
entries equal to zero except \(\tilde{\bm{b}}^y_{k,J+1} = 1\), and,
for \(j(\bar{O}_{(k-1)})\le J-1\),
\(\tilde{\bm{b}}^y_{k,j(\bar{O}_{(k-1)})+1} = 1\) and
\(\tilde{\bm{b}}^y_{k,j(\bar{O}_{(k-1)})} = -1\).

\subsection{Highly adaptive lasso estimation}
\label{sec:highly:adaptive:lasso}

As previously presented, we consider intensities on the form
\begin{align*}
  \Lambda^{\cdot} (dt \mid \F_{t-}) &=  \lambda^{\cdot} (t \mid \F_{t-}) dt =
                                      \1\lbrace  T^{\mathrm{end}} \ge t\rbrace
                                \exp ( f^{\cdot}(t, \rho(\bar{N}^y(t-)), A,L)) dt ,
\end{align*}
with the only difference compared to \eqref{eq:ag:multi:models} being
that we have collapsed the baseline hazard from the models presented
into \(f^{\cdot}\). As noted, then examples of \(\rho\) include
dependence on \(\bar{N}^y(t-) = (N^y(s) \,:\, s < t )\) through, for
example, \(\1\lbrace N^y(t-) \ge 1\rbrace\) (any jumps in the past),
or \(N^y(t-)\) (number of jumps in the past).
How the dependence plays out, is further determined by
\(f^{\cdot}\). In this section we will use notation as follows:
\begin{align*}
  X &=(L, A) \in\mathcal{X}\subset\R^{d+1}, \quad
  V =\rho(\bar{N}^y(t-)) \in\mathcal{V}\subset\R^{d'}. 
\end{align*}
We then define partitionings \(\mathcal{X} = \cupdot_m \mathcal{X}_m\)
and \(\mathcal{V} = \cupdot_m \mathcal{V}_m\) into cubes
\(\mathcal{X}_m\) and \(\mathcal{V}_m\), and let
\(x_m \in\mathcal{X}_m\) and \(v_m \in \mathcal{V}_m\) denote left
endpoints of the cubes \(\mathcal{X}_m\) and \(\mathcal{V}_m\),
respectively.
We further introduce indicator basis functions:
\(\phi_{r}(t) = \1 \lbrace t_{r} \le t \rbrace\);
\( \phi_{m,\mathcal{S}} (x ) = \1 \lbrace x_{m,\mathcal{S}} \le
x_{\mathcal{S}} \rbrace\) for a subset of indices
\( \mathcal{S} \subset \lbrace 1,\ldots, d+1\rbrace\); and
\( \phi'_{m,\mathcal{S}} (v ) = \1 \lbrace v_{m,\mathcal{S}} \le
v_{\mathcal{S}} \rbrace\) for a subset of indices
\( \mathcal{S} \subset \lbrace 1,\ldots, d'\rbrace\). Here
\(x_{m,\mathcal{S}}\) and \(x_{\mathcal{S}}\) denote
\(\mathcal{S}\)-specific coordinates of \(x_m\) and \(x\),
respectively, and \(v_{m,\mathcal{S}}\) and \(v_{\mathcal{S}}\)
likewise denote \(\mathcal{S}\)-specific coordinates of \(v_m\) and
\(v\).
Consider for example the case that
\(V=\rho(\bar{N}^y(t-))= N^y(t-)\in\mathcal{V} \subset\lbrace
0,1,\ldots \rbrace\).  Then \(\mathcal{V} = \cupdot_m [v_m, v_m+1) \)
defines a partitioning into disjoint intervals
\(\mathcal{V}_m = [v_m,v_m+1)\), and the corresponding indicator basis
functions are \(\phi'_{m} (v ) = \1 \lbrace v_{m} \le v \rbrace\).

Denoting by \( \sum_{\mathcal{S}}\) the sum of all subsets
\( {\mathcal{S}}\) of indices, and by \( \sum_{m}\) the sum over
partitionings, the approximation \( f^{\cdot}_{\beta}\) of
\(f^{\cdot}\) with support over the endpoints of the partitionings now
admits a representation over indicator basis functions as follows
\begin{align}
  f^{\cdot}_{\beta}(t,v,x) =  \sum_{r=0}^{R} \phi_{r}(t) \beta_{r}
 + \sum_{r=0}^{R}
                   \sum_{\mathcal{S}\subset \lbrace 1,\ldots, d+1\rbrace}
                   \sum_{m}  \phi_{r}(t)\phi_{m,\mathcal{S}} (x) \beta_{r,m,\mathcal{S}} \\
 \qquad + \, \sum_{r=0}^{R}
                   {\sum_{\mathcal{S}\subset \lbrace 1,\ldots, d'\rbrace}
		   \sum_{m}}
		   \phi_{r}(t)\phi'_{m,\mathcal{S}} (v) \beta'_{r,m,\mathcal{S}} \label{eq:hal:main:N:y}\\
 \qquad + \, \sum_{r=0}^{R}
                  \sum_{\mathcal{S}\subset \lbrace 1,\ldots, d+1\rbrace}  \sum_{\mathcal{S}'\subset \lbrace 1,\ldots, d'\rbrace}
                   \sum_{m}\sum_{m' }  \phi_{r}(t)\phi_{m,\mathcal{S}} (x)\phi'_{m',\mathcal{S}'} (v) \beta_{r,m,\mathcal{S},m',\mathcal{S}'} \label{eq:hal:all:interactions}. 
\end{align}
Here, the \(\beta\) coefficients represent pointmass assigned by
\(f_{\beta}^{\cdot}\) to its support points.
One may note that the dependence of \(f^{\cdot}_{\beta}\) on
\(V= \rho(\bar{N}^y(t-))\) plays out through \eqref{eq:hal:main:N:y}
(main effects, and interactions with time) and
\eqref{eq:hal:all:interactions} (all interactions), and that there are
only finitely many ways in which \(f^{\cdot}_{\beta}\) can depend on
the history of the recurrent event process (even when \(V\), or part
of \(V\), is continuous). This all fits into the estimation and
implementation framework described in Section
\ref{sec:clever:covar:g:comp}.
The sectional variation norm
\citep{gill1995inefficient,van2017generally} of \( f^{\cdot}_{\beta}\)
becomes a sum over the absolute values of its coefficients, i.e., it
equals the \(L_1\)-norm of the coefficient vector. We can now define
the highly adaptive lasso estimator as an empirical risk minimizer
\(\hat{f}^{\cdot}_{n}=f^{\cdot}_{\hat{\beta}^{\cdot}_n}\), defined as
\(\hat{\beta}^{\cdot}_n = \underset{ \beta}{\argmin} \,\, \mathbb{P}_n
\mathscr{L}(f^{\cdot}_\beta)\), s.t.
\( \Vert \beta \Vert_1 \le \mathscr{M}\), with the log-likelihood loss
\( \mathscr{L}(\Lambda^{\cdot} ) (O) = -\int_0^{\tau} \log
\lambda^{\cdot}(t \mid \F_{t-}) N^{\cdot}(dt) + \int_0^{\tau}
\Lambda^{\cdot} (dt \mid \F_{t-})\).  Following
\cite{rytgaard2021estimation}, this can be implemented as an
\(L_1\)-penalized Poisson regression with the indicator functions
\(\phi_{r}(t)\), \(\phi_{r}(t)\phi_{\mathcal{S},m} (x) \),
\(\phi_{r}(t)\phi'_{\mathcal{S}',m'} (v)\) and
\( \phi_{r}(t)\phi_{\mathcal{S},m} (x)\phi'_{\mathcal{S}',m'} (v)\) as
covariates and
\(\beta_{r}, \beta_{r,\mathcal{S},m},
\beta'_{r,\mathcal{S}',m'},\beta_{r,\mathcal{S},m,\mathcal{S}',m'}\)
as the corresponding coefficients.

\section{Simulation study}
\label{sec:simulation:study}

We consider a simulation study to demonstrate the proposed methodology
and as a proof of concept. We explore a setting in which all intensity
processes are allowed to depend on the recurrent event process in a
simple manner, and compare our TMLE estimator to an unadjusted
nonparametric estimator, as well as a simplified version of TMLE
including only baseline information (with simple plug-in estimation of
the target parameter and of clever covariates, c.f., Section
\ref{rem:recurrent:events:no:dependence}).


Let \(L_1 \sim \mathrm{Unif}(-1,1)\), \(L_2 \sim \mathrm{Unif}(0,1)\)
and \(L_3 \sim \mathrm{Unif}(0,1)\), and let
\(A\in\lbrace 0,1\rbrace\) be randomized with \(P(A=1)=1-P(A=0)=0.5\).
Recurrent events are simulated according to the multiplicative
intensity model
\begin{align}
  \lambda^y(t \mid \F_{t-}) = \lambda_{\mathrm{baseline}}^y (t) \exp( \alpha^y + \beta^y_A A  + \beta^y_{L_1} L_1^2 + \beta_{N^y}^y
  \1\lbrace N^y(t-) \ge 1 \rbrace ),
  \label{eq:sim:model:y}
\end{align}
where \(\beta^y_{N^y}\) controls the dependence on whether there were
any past recurrent events. Likewise, censoring events are simulated
according to the intensity model
\begin{align}
  \lambda^c(t \mid \F_{t-}) = \lambda_{\mathrm{baseline}}^c (t) \exp( \alpha^c + \beta^c_A A  + \beta^c_{L_1} L_1  + \beta_{N^y}^c
  \1\lbrace N^y(t-) \ge 1 \rbrace ),
  \label{eq:sim:model:c}
\end{align}
where, like above, \(\beta^c_{N^y}\) controls the dependence on
whether there were any past recurrent events. Finally, failure events
are simulated according to the intensity model
\begin{align}
  \lambda^d(t \mid \F_{t-}) = \lambda_{\mathrm{baseline}}^d (t) \exp( \alpha^d +  \beta^d_A A + \beta^d_{L_1} L_1  + \beta_{N^y}^d
  \1\lbrace N^y(t-) \ge 1 \rbrace ),
  \label{eq:sim:model:d}
\end{align}
again with \(\beta^d_{N^y}\) controlling the dependence on whether
there were any past recurrent events.  The coefficient values
\(\beta^y_A = 1.2\), \(\beta^d_A =0.7\) and \(\beta^c_A =0\) are fixed
throughout, and the baseline intensities
\(\lambda_{\mathrm{baseline}}^y\), \(\lambda_{\mathrm{baseline}}^c\),
and \(\lambda_{\mathrm{baseline}}^d\) are generated from Weibull
distributions. Our main focus is on a primary simulation setting with
\(\beta^y_{N^y}=2.1\), \(\beta^d_{N^y}=1.4\) and
\(\beta^c_{N^y}=1.8\), and \(\beta^y_{L_1} = 2.1 \),
\(\beta^d_{L_1} =0.7 \) and \(\beta^c_{L_1} = 1.4 \), and a secondary
simulation setting with independent censoring, i.e.,
\(\beta^c_{N^y} = \beta^d_{L_1} = 0 \). In both these settings, we
further consider heterogeneity arising only from dependence on the
past of the recurrent event process, corresponding to the coefficient
values \(\beta^y_{L_1} = \beta^d_{L_1} = \beta^c_{L_1} = 0 \). We vary
\(\alpha^c \in \R\) to control the amount of censoring.
Further variations over combinations of
\(\beta^y_{N^y}, \beta^d_{N^y}, \beta^c_{N^y} , \beta^y_{L_1},
\beta^d_{L_1}, \beta^c_{L_1} \) are considered in the Supplementary
Material.

We consider estimation of the target parameter
\(\Psi(P) = \EE_{P^{G^*}} [ N^y(\tau) ]\), with \(\tau = 1.2\), under
the simple intervention
\(\pi^*(a \mid \ell) = \1 \lbrace a= 1\rbrace\), i.e., the target
parameter represents the treatment-specific expected number of
recurrent events by time \(\tau\).


We consider different estimators for the target parameter as
follows:
\textit{1. Unadjusted nonparametric estimator.}  With
\(\hat{S}^d_n (\cdot \mid A=1)\) denoting the (stratified)
Kaplan-Meier estimator, and \(\hat{\Lambda}_n^y(\cdot \mid A=1)\)
denoting the (stratified) Nelson-Aalen estimator, an unadjusted
nonparametric estimator for the target parameter is computed as
\( \hat{\psi}^{\mathrm{snp}}_n = \int_{0}^{\tau} \hat{S}^d_n (t- \mid
A=1) \hat{\Lambda}_n^y(dt \mid A=1) \).
   \textit{2. Working model based TMLE}. With initial estimators based
   on models for the intensities that leave out the dependence on
   \( N^y(t-)\)
   , a working model based
   TMLE estimator is constructed with a targeting step carried out
   using estimators for the clever covariates as presented in Section
   \ref{rem:recurrent:events:no:dependence} and by plugging the
   updated estimators into the formula displayed in
   \eqref{eq:simple:nonparametric}.
   \textit{3. TMLE}. With initial estimators based on correctly
   specified models for all intensities
   , 
   and the targeting step is carried out as outlined in Section
   \ref{sec:special:tmle}.
\textit{4. HAL-TMLE}. The HAL-TMLE estimator is constructed as the
  TMLE estimator, only based on HAL estimators for all
  intensities. The HAL estimators are here implemented in a
  simplified version, including (with our notation from Section
  \ref{sec:highly:adaptive:lasso}) \(X= (L_1,L_2,L_3,A)\) and
  \(V = \1\lbrace N^y(t-) \ge 1 \rbrace\). 
  
In the Supplementary Material, to illustrate double robustness, we
also consider misspecified models for 2. and 3. that include the
linear effect of \(L_1\) rather than \(L_1^2\) when fitting the model
for the intensity of the distribution of recurrent events. 

\begin{figure}[!ht]
\centering

\begin{subfigure}[t]{1\textwidth}
  \makebox[\textwidth][c]{\includegraphics[width=1.0\textwidth]{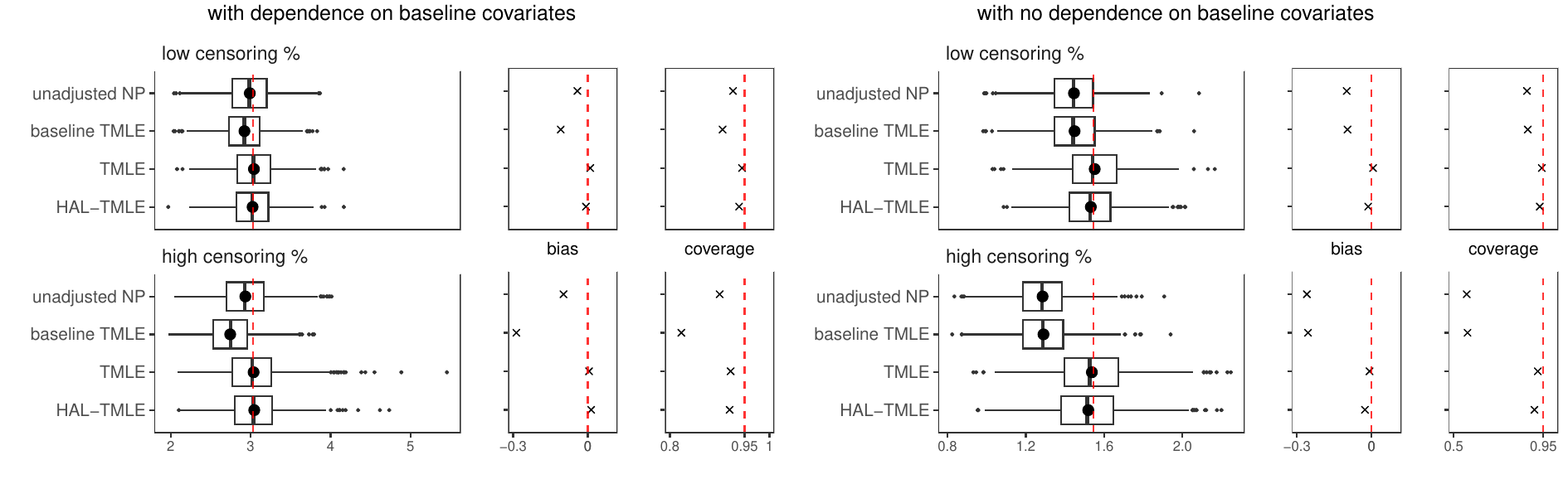}}
    \vspace{-1.5cm}
  \caption{Simulation results from the primary simulation setting
    (with dependent censoring).}\label{fig:1}
\end{subfigure}

\begin{subfigure}[t]{1\textwidth}
  \makebox[\textwidth][c]{\includegraphics[width=1.0\textwidth]{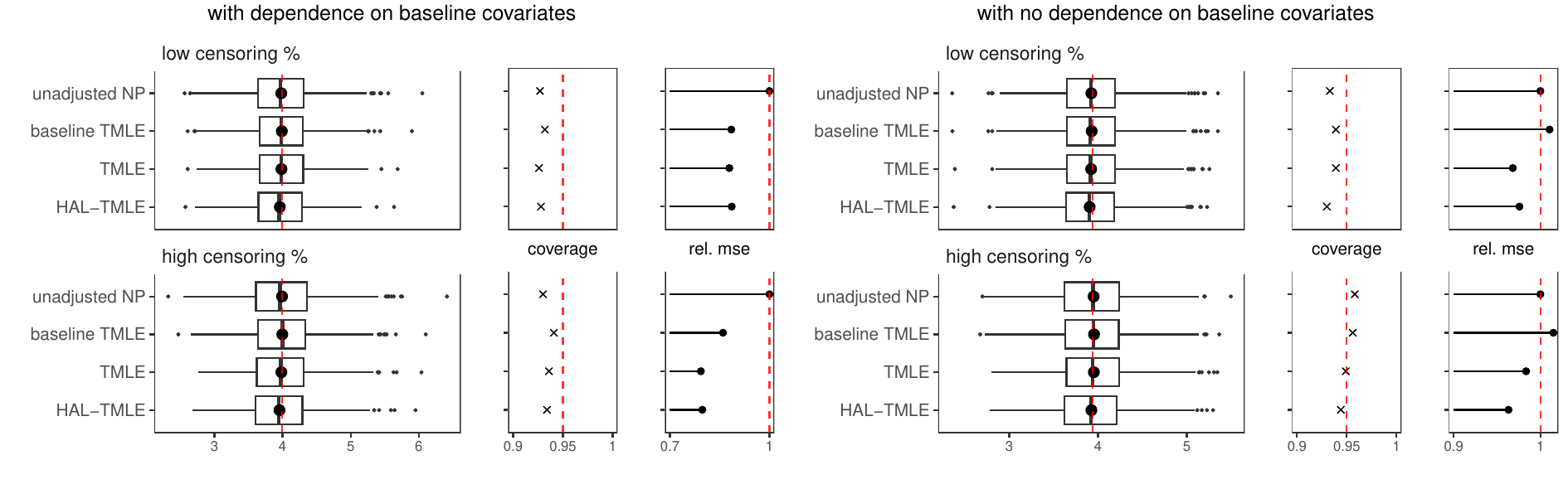}}
  \vspace{-1.5cm}
  \caption{ Simulation results from the setting with independent
    censoring (i.e., \(\beta^c_{N^y} = \beta^d_{L_1} = 0
    \)).  }\label{fig:2}
\end{subfigure}

\caption{Simulation results from the setting with dependent censoring
  (panel (a)) and independent censoring (panel (b)). In both panels,
  the left column shows results from the setting with a predictive
  covariate, and the right column results from the setting where there
  is no dependence on baseline covariates; moreover, the upper row of
  each panel shows results from a low censoring percentage setting,
  and the lower row results from high censoring percentage settings.
}\label{fig:12}
\end{figure}

Figure \ref{fig:1} shows the results from the primary simulation
setting with dependent censoring.  We verify that our TMLE estimators
provide unbiased estimation and appropriate coverage across all
variations of low/high censoring and predictiveness of covariates. In
contrast, as a result of ignoring follow-up information on the
recurrent event process, the working model based TMLE is highly biased across all
settings. When heterogeneity arises solely from the dependence on the
past of the recurrent event process, we note that the working model based TMLE
and the unadjusted nonparametric estimator exhibit very similar
performances. The HAL-TMLE does not rely on correct model
specification, and demonstrates comparable precision to the correctly
specified TMLE.

Figure \ref{fig:2} shows the results from the secondary setting with
independent censoring.  Here the unadjusted nonparametric estimator
yields unbiased estimation, as does the working model based TMLE since the
censoring model is correctly specified.
Efficiency, as measured by mean squared error, improves in the
covariate-dependent setting when using the working model based TMLE, which
accounts for predictive covariates, whereas there are no advantages of
the working model based TMLE compared to the unadjusted nonparametric estimator
in the covariate independent setting.
The mean squared error of the (general) TMLE is always better than
that of the working model based TMLE, from taking into account the dependence on
the past of the recurrent event process.
The HAL-TMLE, again, demonstrates comparable performance to the TMLE
relying on correctly specified models, with no model specification.

\section{Real data example}
\label{sec:data:analysis}

We illustrate our methods with the dataset \verb+readmission+ obtained
from the \verb+frailtypack+ \citep{rondeau2012frailtypack} package in
\verb+R+.  These are data from an observational study on readmission
(rehospitalization) of 403 patients with colorectal cancer
\citep[see,][]{gonzalez2005sex}, followed from day of surgery for up
to 6 years (2176 days). There were 109 deaths, 458 (recurrent)
readmission events, and 294 were censored (comprised by migration,
change of hospital, or administrative censoring due to different dates
of surgery). 
Several readmissions can occur for the same
patient, and, as noted in the original analysis by
\cite{gonzalez2005sex}, these events are potentially highly dependent,
which should be accommodated by the estimation methods for reasons as
highlighted in previous sections: 1) if the censoring mechanism
depends on the history of the recurrent event process, any method that
does not take the general dependence into account is potentially
biased, and 2) when the censoring mechanism does not depend on the
history of the recurrent event process, methods that do not take the
general dependence into account are not efficient. 

Based on the data, interest is in assessing the effect of receiving
chemotherapy or not, controlling for the available baseline variables:
sex, tumour stage, and Charlson's index. We consider the
intervention-specific mean outcomes of the number of recurrent events
by time \(\tau = 5\) years (corresponding to 1825 days) under
treatment with chemotherapy and no treatment with chemotherapy. The
difference between these two intervention-specific mean outcomes
represents how many events are on average avoided by the treatment.

In our analysis, we consider (using the same nomenclature as in
Section \ref{sec:simulation:study}) 1) an unadjusted nonparametric
estimator, 2) a working model based TMLE estimator, and 3) a HAL-TMLE
estimator.  For the HAL estimator, we allowed dependence on
\( N^y(t-)\), covariates and chemotherapy through main terms and
two-way interactions.  We emphasize that the HAL-TMLE estimate is all
we are after in the end, and that the other estimates are included
solely for reference. Results can be found in Table
\ref{tab:data:analysis:results}.

\begin{table}[!ht]
  \centering
  \vspace{-0.25cm}
\begin{tabular}{lllllll}
 & \(\hat{\psi}^1_n\) & \(\hat{\sigma}_n^1\) & 95\% CI & \(\hat{\psi}^0_n\) & \(\hat{\sigma}_n^0\) & 95\% CI \\ 
  \toprule\\[-1.1cm]
 unadjusted np & 1.084 & 0.139 & [0.812,1.356] & 1.699 & 0.209 & [1.289,2.109] \\[-0.2cm] 
  working model based TMLE & 1.129 & 0.163 & [0.81,1.448] & 1.662 & 0.204 & [1.262,2.062] \\[-0.2cm]
  HAL-TMLE &  1.18 & 0.169 & [0.849,1.511] & 1.803 & 0.202 & [1.407,2.199] \\ 
  \bottomrule
\end{tabular}
\caption{Results from the data analysis. Shown are the estimated
  treatment-specific (\(\hat{\psi}^1_n\)) and control-specific
  (\(\hat{\psi}^0_n\)) expected number of recurrent events at 5 years
  of follow-up
  .}\vspace{-0.3cm} \label{tab:data:analysis:results}
\end{table}

\section{Discussion}
\label{sec:discussion}

This work presents targeted learning methodology to estimate the
treatment-specific expected number of recurrent events in recurrent
events settings with baseline confounding, informative
right-censoring, competing risks, and dependence on the recurrent
event process. We have presented the efficient influence curve for the
estimation problem, as well as its double robust properties, and
proposed an estimation procedure based on targeted maximum likelihood
estimation with direct plug-in of targeted intensity estimators
solving the efficient influence curve equation. Acknowledging the
challenges of model misspecification, we have highlighted the
incorporation of flexible highly adaptive lasso estimation, allowing
us to learn about dependencies in the data without imposing rigid
functional forms or interaction assumptions.

It is important to clarify that our sequential procedure is not the
same as the longitudinal sequential regression approach, that
sequentially performs regression steps across time-points to avoid
density estimation
\cite{bang2005doubly,van2012targeted,ltmleRpackage}. Instead, our
sequential procedure serves the purpose of directly evaluating the
g-computation formula for a given density fit. This is similar to the
ideas of the previous work on targeted maximum likelihood estimation
in discrete-time longitudinal settings \citep{stitelman2012general},
but with subtle differences in how targeting is carried out; while
targeting in discrete-time settings can be handled one time-point by
one, this is hindered by events happening in continuous time. 
A key strength of our sequential procedure is its explicit evaluation
of the g-computation formula. A limitation, on the other hand, is its
relatively slow performance, although we note that the process could
straightforwardly be parallelized across subjects to reduce
computation time.
There are other promising opportunities for future implementations. To
estimate the target parameter, one may for example proceed with monte
carlo simulations based on the estimators for all the likelihood
factors.  Alternatively, one could consider using the highly adaptive
lasso estimator for all intensities, and evaluate the g-computation
formula along the lines of Section \ref{sec:g:comp:::} without
targeting. General theory on the highly adaptive lasso estimator shows
that such an estimator can achieve asymptotic efficiency under
undersmoothing so that it solves the efficient influence curve
equation \citep{van2019causal}. It will be of interest to look into
whether such an estimator, by leaving out iterative rounds of
targeting update steps, may simplify implementation complexity.

In recurrent events settings with presence of death, such as we have
considered in this work, it is a well-acknowledged challenge that
examining treatment effects only on the recurrent event process can
give a misleading impression of the preventative (or causative) nature
of the treatment, especially when death rates are high. This is simply
due to the fact that individuals can experience recurrent events only
until they encounter the terminal outcome, so that in extreme
situations a treatment found to lower the expected number of recurrent
events may in fact only do so by increasing mortality. To gain full
insights into the effect of a given treatment, estimation could be
targeted for multivariate parameters that assess effects jointly on
both the expected number of recurrent events and survival
probabilities. 

\if0\blind\section*{Funding}

This research is partially funded by the Novo Nordisk Foundation grant
NNF23OC0084961. Additional funding was provided by a philanthropic
gift from Novo Nordisk. \fi


\newpage

\bibliographystyle{chicago}

\end{document}